\documentclass[]{spie}  

\usepackage[]{graphicx}

\title{Polarization Sensitive Multi-Chroic MKIDs} 

\author{
Bradley R. Johnson\supit{a},
Daniel Flanigan\supit{a},
Maximilian H. Abitbol\supit{a},
Peter A. R. Ade\supit{b},\\
Sean Bryan\supit{c},
Hsiao-Mei Cho\supit{g},
Rahul Datta\supit{e}, 
Peter Day\supit{f},
Simon Doyle\supit{b},\\ 
Kent Irwin\supit{d,g}, 
Glenn Jones\supit{a}, 
Sarah Kernasovskiy\supit{d}, 
Dale Li\supit{g},
Phil Mauskopf\supit{c},\\ 
Heather McCarrick\supit{a}, 
Jeff McMahon\supit{e}, 
Amber Miller\supit{a}, 
Giampaolo Pisano\supit{b},\\
Yanru Song\supit{d}, 
Harshad Surdi\supit{c}, and 
Carole Tucker\supit{b}
\skiplinehalf
\supit{a}Department of Physics, Columbia University, New York, NY, 10027, USA; \\
\supit{b}School of Physics \& Astronomy, Cardiff University, Cardiff, CF243AA, UK; \\
\supit{c}School of Earth and Space Exploration, Arizona State University, Tempe, AZ, 85287, USA; \\
\supit{d}Department of Physics, Stanford University, Stanford, CA, 94305-4085, USA; \\
\supit{e}Department of Physics, University of Michigan, Ann Arbor, MI, 48103, USA; \\
\supit{f}NASA, Jet Propulsion Lab, Pasadena, CA, 91109, USA; \\
\supit{g}SLAC National Accelerator Laboratory, Menlo Park, CA 94025, USA \\
}

\authorinfo{E-mail: bjohnson@phys.columbia.edu}

\begin{document} 
\maketitle 


\begin{abstract}
We report on the development of scalable prototype microwave kinetic
inductance detector (MKID) arrays tailored for future multi-kilo-pixel
experiments that are designed to simultaneously characterize the
polarization properties of both the cosmic microwave background (CMB)
and Galactic dust emission.
These modular arrays are composed of horn-coupled,
polarization-sensitive MKIDs, and each pixel has four detectors: two
polarizations in two spectral bands between 125 and 280~GHz.
A horn is used to feed each array element, and a planar orthomode
transducer, composed of two waveguide probe pairs, separates the
incoming light into two linear polarizations.
Diplexers composed of resonant-stub band-pass filters separate the
radiation into 125~to~170~GHz and 190~to~280~GHz pass bands.
The millimeter-wave power is ultimately coupled to a hybrid co-planar
waveguide microwave kinetic inductance detector using a novel,
broadband circuit developed by our collaboration.
Electromagnetic simulations show the expected absorption efficiency of
the detector is approximately 90\%.
Array fabrication will begin in the summer of 2016.
\end{abstract}



\section{INTRODUCTION}
\label{sec:introduction}


Microwave kinetic inductance detectors (MKIDs) are superconducting
thin-film, GHz resonators that are designed to also be optimal photon
absorbers\cite{zmu}.
Absorbed photons with energies greater than the superconducting gap
($\nu > 2 \Delta/h \cong 74~\mbox{GHz} \times (T_c/1~\mbox{K})$) break
Cooper pairs, changing the density of quasiparticles in the device.
The quasiparticle density affects the kinetic inductance and the
dissipation of the superconducting film, so a changing optical signal
will cause the resonant frequency and internal quality factor of the
resonator to shift.
These changes in the properties of the resonator can be detected as
changes in the amplitude and phase of a probe tone that drives the
resonator at its resonant frequency.
This detector technology is particularly well-suited for sub-kelvin,
kilo-pixel detector arrays because each detector element can be
dimensioned to have a unique resonant frequency, and the probe tones
for hundreds to thousands of detectors can be carried into and out of
the cryostat on a single pair of coaxial cables.

\begin{figure}[!t]
\centering
\includegraphics[width=0.95\textwidth]{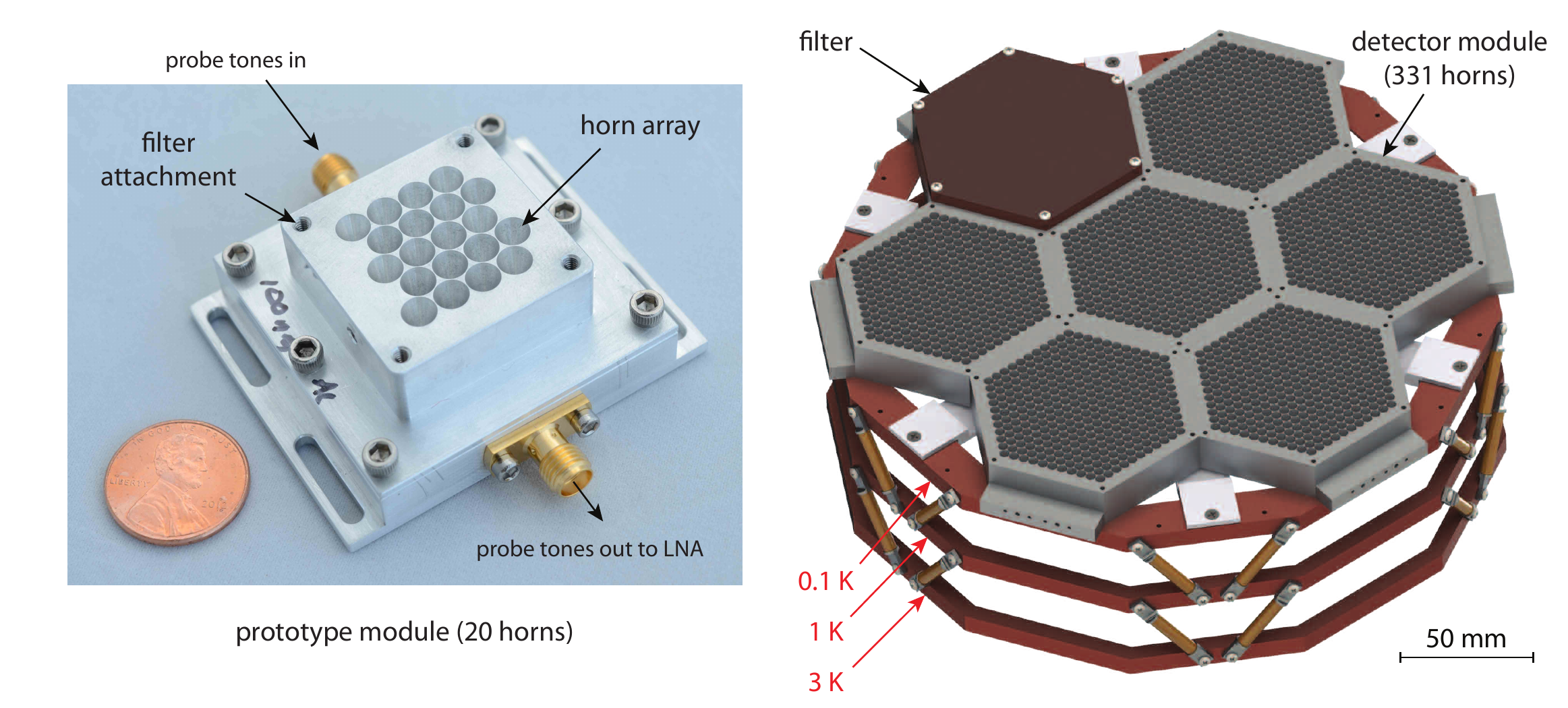}
\caption{
\textbf{Left:} A scalable 20-element prototype module for
dual-polarization LEKIDs.
The module we are currently building for our prototype multi-chroic
MKID arrays looks similar.
\textbf{Right:} Future focal plane concept.
By design, the module architecture shown on the left is scalable to
one of the seven modules shown on the right.
The concept detector array on the right includes 9268 single
polarization detectors spread over two spectral bands.
}
\label{fig:focal_plane_concept}
\end{figure}

In this paper, we report on the development of modular arrays of
horn-coupled, polarization-sensitive MKIDs that are each sensitive to
two spectral bands between 125 and 280~GHz.
The scalable prototype MKID arrays we are developing are tailored for
future multi-kilo-pixel experiments that are designed to
simultaneously characterize the polarization properties of both the
cosmic microwave background (CMB) and Galactic dust emission.
Our device design builds from successful transition edge sensor (TES)
bolometer architectures that have been developed by the Truce
Collaboration\footnote{http://casa.colorado.edu/$\sim$henninjw/TRUCE/TRUCE.html}
and demonstrated to work in receivers on the ACT and SPT
telescopes\cite{thornton2016,actpol,austermann2012,sptpol}.
Detector modules like these could be a strong candidate for a future
CMB satellite mission and/or CMB-S4\cite{bock_2009,cmbs4} because
these future multi-kilo-pixel programs will require efficient
multiplexing schemes and MKID arrays could out-perform current
technologies in this regard (see
Figure~\ref{fig:focal_plane_concept}).


A range of MKID-based instruments have already shown that MKIDs work
well at millimeter and sub-millimeter wavelengths.
Early MKIDs used antenna coupling\cite{Day2006}, and these
antenna-coupled MKIDs were demonstrated at the Caltech Submillimeter
Observatory (CSO) in 2007\cite{Schlaerth2008} leading to the
development of MUSIC, a multi-chroic antenna-coupled MKID
camera\cite{golwala+12}.
A simpler device design that uses the inductor in a single-layer $LC$
resonator to directly absorb the millimeter and sub-millimeter-wave
radiation was published in 2008\cite{doyle}.
This style of MKID, called the lumped-element kinetic inductance
detector (LEKID), was first demonstrated in 2011 in the 224-pixel NIKA
dual-band millimeter-wave camera on the IRAM 30~m telescope in
Spain\cite{monfardini}.

Laboratory studies have shown that state-of-the-art MKID and LEKID
designs can achieve photon noise limited
performance\cite{mauskopf14,McKenney2012}.
Photon noise limited horn-coupled LEKIDs sensitive to 1.2~THz were
recently demonstrated\cite{Hubmayr2014} and these detectors will be
used in BLAST-TNG\cite{galitzki2014,dober2014}.
Members of our collaboration conducted studies of horn-coupled LEKIDs
to see if LEKIDs would be suitable for CMB polarimetry.
These studies revealed that the sensitivity of the LEKID variety of
MKID can be compared with that of state-of-the-art TES
bolometers\cite{flanigan_2016,mccarrick_2014}.
On-chip spectrometers based on MKIDs are currently being
developed\cite{superspec2012,microspec2013}.
And a large format sub-millimeter wavelength camera, called A-MKID,
with more than 10,000 pixels and readout multiplexing factors greater
than 1,000 has been built and is currently being
commissioned\cite{baryshev+2014}.


\begin{figure}[!t]
\centering
\includegraphics[width=0.9\textwidth]{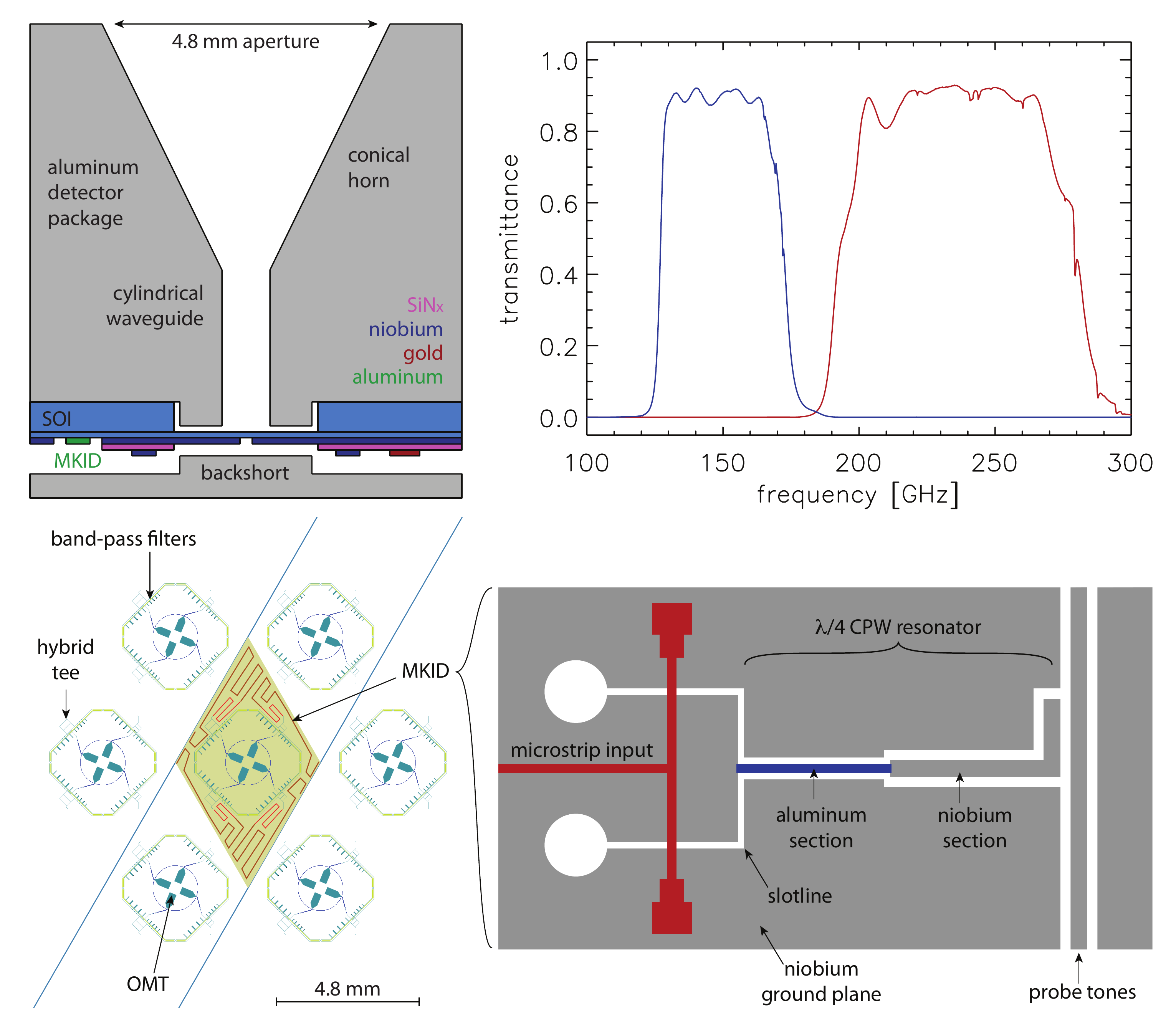}
\caption{
\textbf{Top~Left:} A cross-sectional view of one focal-plane element.
In an effort to minimize two-level system (TLS) noise, the MKID
sensing element is deposited directly on the silicon wafer, and it is
not covered with silicon nitride.
\textbf{Bottom~Left:} A scale drawing of the dual-polarization
multi-chroic MKID device we propose to develop.
\textbf{Bottom~Right:} A schematic of the co-planar waveguide MKID we
are developing.
Photons from the sky are brought to the detector on a microstrip from
the hybrid tee.
These photons then couple to our resonator and are absorbed in the
aluminum section of the CPW $\lambda/4$ resonator.
\textbf{Top~Right:} End-to-end electromagnetic simulations show the
expected absorption efficiency is approximately 90\% across the
150~GHz and the 235~GHz spectral bands.
}
\label{fig:pixel_design}
\end{figure}


\section{METHODS}
\label{sec:methods}


Our horn-coupled, multi-chroic devices are based on the polarimeters
that were developed for the Advanced ACTPol
experiment\cite{henderson2016,MCdetectorsDatta}.
These devices were recently deployed and field tested, and early
indications are they work well.
Our MKID development program is focused on re-optimizing this design
for silicon-on-insulator (SOI) wafers and replacing the TES bolometers
with hybrid co-planar waveguide (CPW) MKIDs.
Our design is shown in Figure~\ref{fig:pixel_design}.
The microstrip to CPW coupling technology is a critical part of our
development program, and the move to SOI wafers is motivated in
Section~\ref{sec:mkid_design}.
Note that the nominal Advanced ACTPol design uses a ring-loaded
corrugated feed.
For our laboratory development work we will use a conical horn for
simplicity and switch to profiled horns in the future.
Profiled horns are easier to fabricate and they have been shown to
perform like corrugated feeds\cite{zeng_2010}.


\subsection{Horn Coupling and RF Circuit}
\label{sec:horn_rf}


In our prototype design, a conical horn is used to feed each array
element.
Each horn is machined into a monolithic horn plate that also serves as
both the top of the detector module and the mounting surface for the
MKID arrays.
The bottom plate, which closes the module, also contains backshorts,
which are used to optimize photon coupling.
Light emerging from the cylindrical waveguide is coupled to a
broadband orthomode transducer (OMT).
A choke around the exit aperture of the waveguide minimizes lateral
leakage of the fields.
The OMT is composed of two probe pairs, and it separates the incoming
light into two linear polarizations.
For example, one linear polarization couples to one pair of probes,
and the wave then propagates through identical electrical paths in the
subsequent millimeter-wave circuit en route to the MKID absorbing
element.
Along each path, a broadband CPW-to-microstrip transition composed of
seven alternating sections of CPW and microstrip is first used to
transition the radiation onto microstrip lines.
Next, diplexers composed of two separate five-pole resonant-stub
band-pass filters separate the radiation into 125~to~170~GHz and
190~to~280~GHz pass bands.
The signals from opposite probes within a single sub-band are then
combined onto a single microstrip line using the difference output of
a hybrid tee.
Signals at the sum output of the hybrid are routed to a termination
resistor and discarded, while the output of the difference port is
detected.

These polarimeters operate over a 2.25:1 ratio bandwidth over which
cylindrical waveguide becomes multi-moded.
However, the TE11 mode, which has the desirable polarization
properties, couples to opposite fins of the OMT with a 180$^{\circ}$
phase shift while the higher order modes, which also couple
efficiently to the OMT probes have a 0$^{\circ}$ phase shift.
This phase difference allows the hybrid tee to isolate the TE11 signal
at the difference port and reject the unwanted modes at the sum port.
This ensures single-moded performance over our 2.25:1 bandwidth.
The architecture described above offers a frequency
\textit{independent} polarimeter axis defined by the orientation of
the planar OMT.
Figure~\ref{fig:pixel_design} shows a schematic of our
microstrip-to-CPW coupler.
The the HFSS/Sonnet simulation results in this figure show the
expected absorption efficiency of the detector is approximately 90\%
taking into account all of the elements in the circuit except the OMT
probes.


\begin{figure}
\centering
\includegraphics[width=\textwidth]{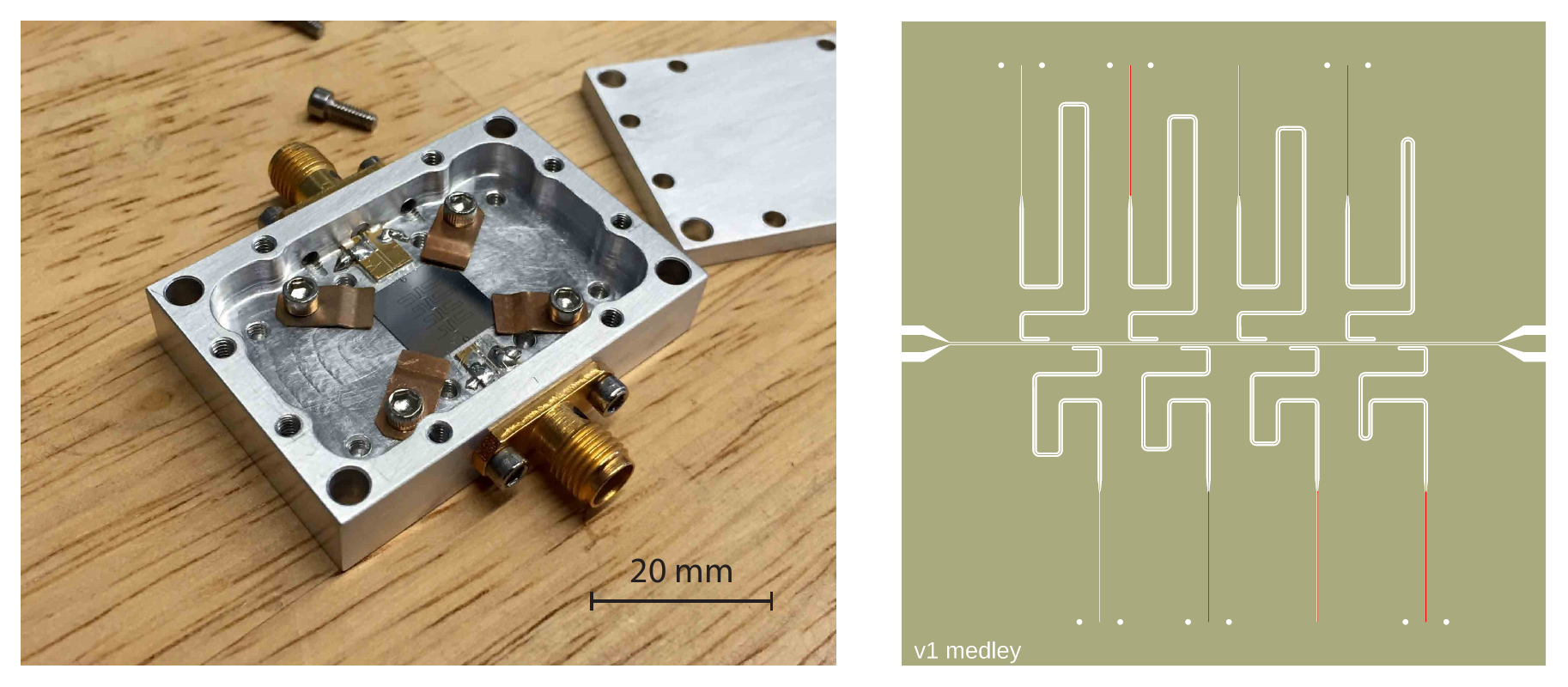}
\caption{
A photograph of our MKID test chip mounted inside an aluminum package
designed for dark testing.
The test chip layout is shown on the right.
Each chip contains eight hybrid CPW MKIDs with different lengths and
meander properties.
The red elements are made from aluminum and the gray regions are 
niobium.
}
\label{fig:prototypes}
\end{figure}


\subsection{MKID Design}
\label{sec:mkid_design}


The total instrument noise is the quadrature sum of the detector noise
and the photon noise, and the fundamental performance goal is to
achieve a sensitivity that is dominated by the random arrival of
background photons.
For an MKID, the detector noise includes contributions from three sources:
generation-recombination (g-r) noise, two-level system (TLS) noise,
and amplifier noise\cite{zmu}.
The g-r noise comes from the random recombination of quasiparticles.
At typical operating temperatures and optical loads, quasiparticle
generation noise is dominated by optical generation -- the photon
noise -- and thermal generation is negligible.
TLS noise is produced by dielectric fluctuations due to quantum two
level systems in amorphous dielectric surface layers surrounding the
MKID.
The scaling of TLS noise with operating temperature, resonator
geometry, and readout tone power and frequency has been extensively
studied experimentally.
We used a semi-empirical model\cite{Gao2008b} to design the
resonators in order to reduce TLS noise.
Finally, the amplifier noise is the electronic noise of the readout
system, which is dominated by the cryogenic microwave low-noise
amplifier.
Our modeling of these noise sources indicates that the detector
sensitivity will be limited by photon noise under the 5--20~pW optical
loads typically present for 150~GHz detectors in ground-based CMB
experiments.

Our MKID design is based on a quarter-wavelength CPW resonator (see
Figure~\ref{fig:pixel_design}~\&~\ref{fig:prototypes}).
The design uses a hybrid CPW transmission line composed of two
different metals.
The ground plane is made of a superconductor with Cooper pair binding
energy greater than the optical photon energy; our design uses
niobium.
Most of the CPW center trace is made from the same high-gap
superconductor, except for a small active region adjacent to the
grounded end.
This active region is made from a lower-gap superconductor in which
the optical photons can break Cooper pairs; our design uses aluminum.
The quasiparticles excited by optical photons alter the dissipation
and kinetic inductance of the device, and these changes are probed by
a tone produced in the ROACH-based readout (see
Section~\ref{sec:readout}).
Devices like these have achieved photon-noise-limited performance over
a wide range of millimeter and sub-millimeter wavelengths with optical
loading levels well under 1~pW\cite{janssen_2014,yates}.

Because of the gap difference, the quasiparticles are trapped in the
active region, the volume of which can be reduced to increase the
device responsivity.
The active region of our baseline design is 2~mm long with an 8~$\mu$m
wide aluminum center line and a 5 to 10~$\mu$m gap.
Greater than 90\% of the millimeter-wave power is dissipated here due
to Cooper pairs breaking.
The length of the second section varies from detector to detector
because it is used to tune the resonant frequency.
This section can have a much wider gap to the ground plane, which
reduces TLS noise.
In our baseline design, the second section of CPW is $\sim$8~mm long
with a 10~$\mu$m niobium center trace and a gap of 30~$\mu$m to the
niobium ground plane.
This $\sim$10~mm length of transmission line will have a resonant
frequency of approximately 3~GHz.

The millimeter-wave power is coupled from the microstrip output of the
hybrid tee to the CPW of the MKID using a novel, broadband circuit
developed by our collaboration\cite{surdi2016} (see
Figure~\ref{fig:pixel_design}).
First, the power is evenly divided in-phase onto two microstrips each
with twice the impedance of the incoming microstrip.
Each branch feeds a standard broadband microstrip-to-slotline
transition, where the slotline is formed in the niobium ground plane
that is common to the microstrip and the MKID CPW.
The two slotlines are then brought together and become the gaps of the
CPW transmission line, efficiently coupling the radiation into the
aluminum CPW center line, where it dissipates by exciting
quasiparticles.
The slotline is electrically short at the resonant frequency of the
MKID, and thus it does not impact the microwave characteristics of the
resonators.

%
%


\subsection{Fabrication}
\label{sec:fabrication}


The arrays will be fabricated on SOI wafers 100~mm in diameter.
Each SOI wafer consist of a 5~$\mu$m thick silicon device layer and a
350~$\mu$m thick handling wafer held together by a 0.5~$\mu$m thick
oxide layer.
The MKID arrays are fabricated on the device layer, which is made of
high-purity, float-zone silicon ($>$10~k$\Omega$\,cm resistivity).
The first metal deposition is a niobium film which is patterned to
produce the ground plane and the OMT.
The aluminum film that will form the MKID sensing element is then
deposited after the niobium ground plane is patterned.
The aluminum MKID element is co-planar with the niobium ground plane
on the 5~$\mu$m thick silicon device layer.
Silicon nitride is then deposited as an electrically insulating
dielectric material, which will be used to form the substrate of the
microstrip lines.
A second niobium film is used to form the top microstrip line.
Our design uses crossunders rather than crossovers, so a second
silicon nitride and third niobium layer are not need.
The layer of silicon nitride is removed to reduce loss and TLS noise.
A gold film is deposited and patterned to construct the termination in
the hybrid tee.
The thick silicon handling wafer and the oxide underneath the OMT and
the MKID will be removed using deep reactive ion etching (DRIE) to
improve the bandwidth and the optical coupling and to minimize TLS
noise.


\begin{figure}[!t]
\centering
\includegraphics[width=\textwidth]{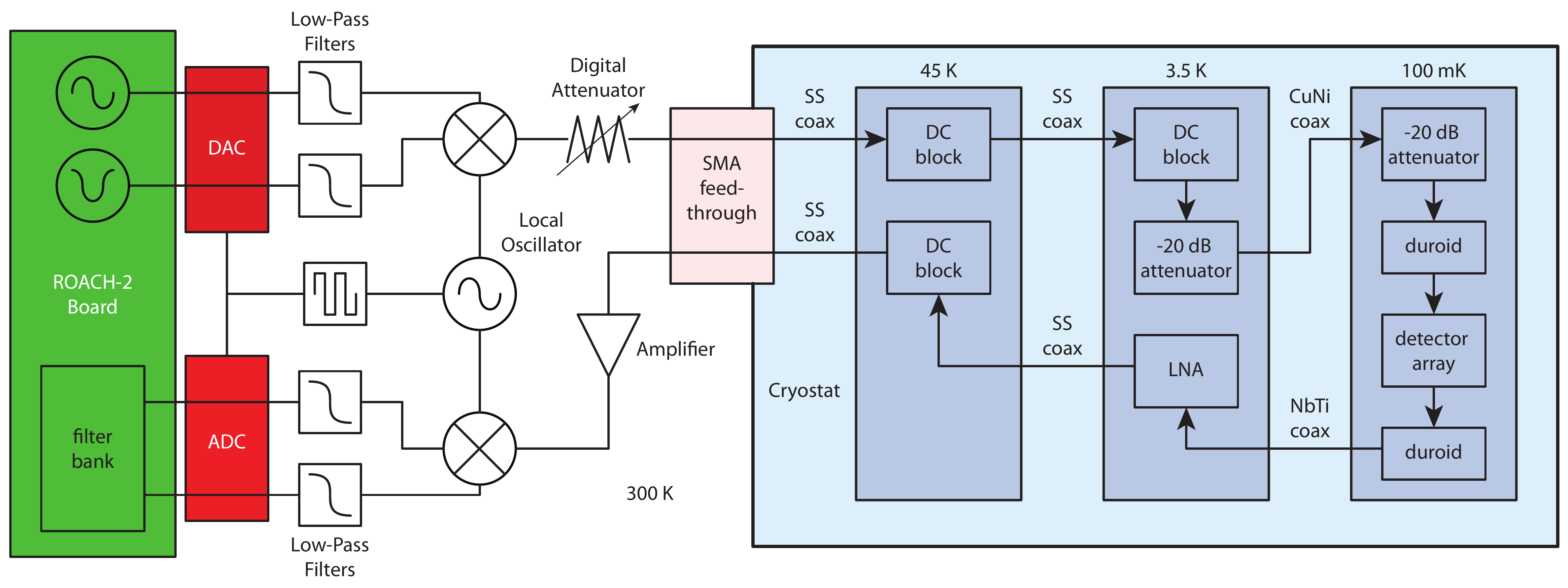}
\includegraphics[width=\textwidth]{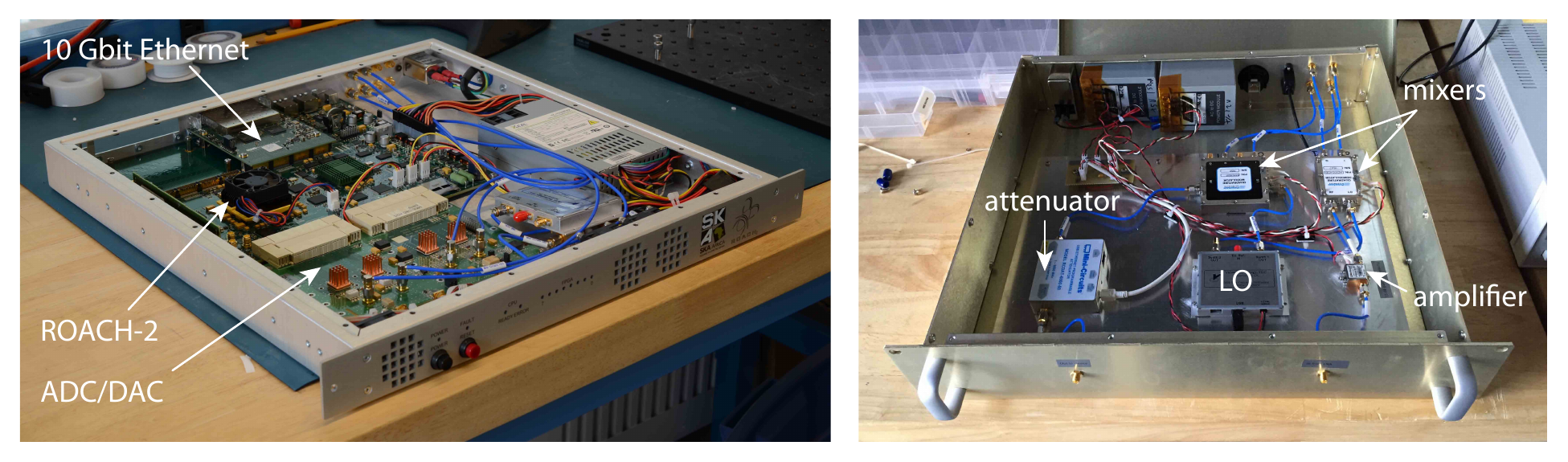}
\caption{
\textbf{Top:} Readout schematic showing the probe tone path.
\textbf{Bottom Left:} The ROACH-2 with the DAC/ADC.
\textbf{Bottom Right:} The analog signal conditioning hardware.  This
chassis houses the filters, room-temperature mixers, attenuator, warm
amplifier, and the local oscillator shown in the schematic above.
}
\label{fig:readout}
\end{figure}


\subsection{Readout}
\label{sec:readout}


All tests of the resonator performance are made by injecting
sinusoidal tones near the resonant frequency of the MKID and measuring
the amplitude and phase of the emerging waveform.
To measure the resonant frequency and quality factor, the frequency of
the sine wave is stepped through the resonance, effectively measuring
the complex forward transmission ($S_{21}$) as with a vector network
analyzer.
Once the resonant frequency has been found from these sweeps, the
probe tone frequency is tuned to this resonant frequency and a complex
voltage time series is recorded.
This time series can then be decomposed into fluctuations of the
resonant frequency and of the quality factor of the resonator.
Standard spectral analysis is then used to determine the noise
characteristics of the detector.

While all of the measurements just described can be done for a single
resonator at a time using a microwave synthesizer and homodyne mixer,
one of the major advantages of MKIDs is that a digital waveform
generator and a digital filter bank and demodulator can be used to
measure hundreds of resonators simultaneously by superimposing sine
waves of different frequencies.
Several such systems have been developed and deployed\cite{duan+10}.
Many of these systems, including one developed at Columbia University
for this work, are based around the CASPER ROACH-1 and ROACH-2
FPGA boards\footnote{https://casper.berkeley.edu/}.
A schematic and photos of a ROACH-based read out system developed at
Columbia are shown in Figure~\ref{fig:readout}.
Using this system, we can characterize hundreds of resonators
simultaneously, so readout of the 80 resonators per prototype detector
module is straightforward.

While the ROACH-1 is a robust and well-proven board, it is now several
years old and is being superseded by the ROACH-2 board.
We have ported our readout design to the ROACH-2, which allows us to
read out over a thousand resonators at full bandwidth for extensive
laboratory testing, and is suitable for deployment of a future large
format array.
We have designed an analog signal conditioning system based around
Polyphase Microwave quadrature modulators and demodulators to convert
the baseband signals generated and analyzed by the ROACH to the target
3--4~GHz readout band.


\begin{figure}[t]
\centering
\includegraphics[width=0.85\textwidth]{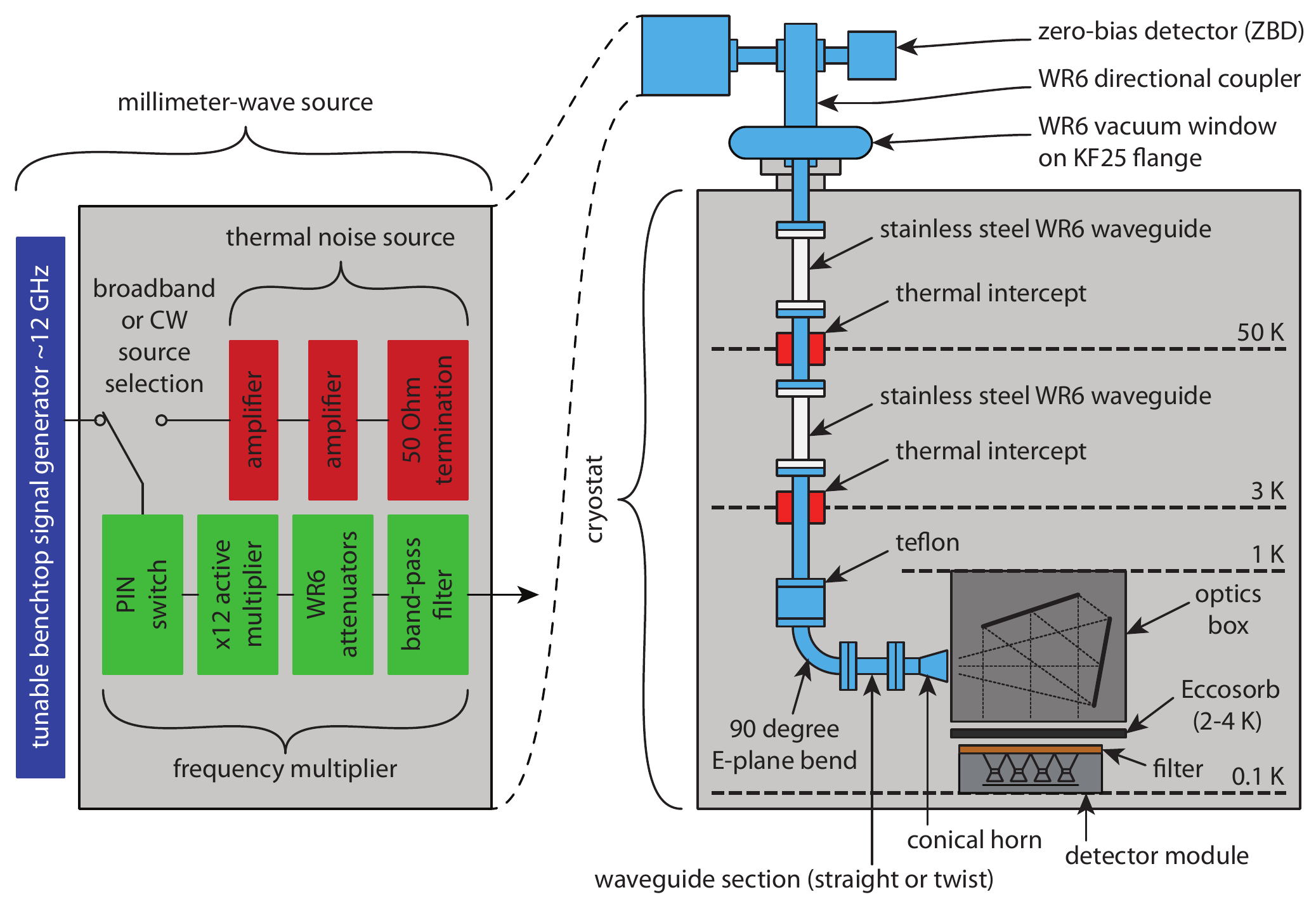}
\caption{
A schematic of the testing apparatus that is already successfully
operating in the laboratory at Columbia.
Millimeter-wave radiation is coupled into the cryostat using WR-6
waveguide, and this radiation illuminates the detector array via a
horn and a collimating reflective optical assembly.
The cold Eccosorb acts as both an attenuator and a diffuse background
signal for the detectors.
By changing the temperature of the Eccosorb we obtain an absolute
brightness temperature calibration.
We will use this existing setup to characterize the spectral response
of the lower frequency band, and a similar $\sim$235~GHz source will
be constructed to test the high-frequency spectral band.
}
\label{fig:testing_schematic}
\end{figure}


\section{TESTING PLAN}
\label{sec:testing_plan}


To characterize the detector arrays, we will install them in
superconducting aluminum detector packages and cool the assemblies to
approximately 120~mK using the STAR~Cryoelectronics DRC-100 cryostat
system.
This system includes a two-stage adiabatic demagnetization
refrigerator (ADR) backed by a Cryomech PT-407 pulse tube cooler.
We have developed and built a cryogenic optical test setup inside this
cryostat, which is schematically described in
Figure~\ref{fig:testing_schematic}.
%
%
The testing system consists of two main photon sources.


\subsection{Millimeter-Wave Source} 
\label{sec:millimeter-wave_source}


A millimeter-wave signal source, which is based on a Millitech active
12-times frequency multiplier, was built by the group at Columbia.
The multiplier can be driven by a sweepable microwave signal
generator, which is useful for measuring the frequency response of the
detectors, or it can be driven by a broadband noise source to mimic a
diffuse astrophysical background signal such as the CMB or Galactic
dust emission.
%
%
The input to the frequency multiplier is connected through a PIN diode
switch, which allows the signals to be chopped on and off with
sub-microsecond time resolution.
This functionality provides a direct way of measuring the response
time of the detectors.
The millimeter-wave source is operated outside the cryostat at room
temperature, and photons are brought to the detectors using WR-6
waveguide.
The waveguide is passed into the cryostat through a vacuum waveguide
window, and sections of stainless steel waveguide are used to decrease
the thermal load on the cryogenic system.
Inside the cryostat, the WR-6 waveguide feeds a horn (mounted at 4 K)
and a small crossed-Dragone optics box (mounted at 1 K), which
converts the diverging horn beam into a plane wave that illuminates
the detectors.
This configuration will be ideal for testing the 150~GHz spectral band
of the detectors.
To test the high frequency band, we will build a similar
frequency multiplier-based source to cover the 190-280~GHz band.

To test the polarization response, we currently have interchangeable
waveguide sections with $0^\circ$, $45^\circ$, and $90^\circ$ twists
to set the polarization angle of the radiation emitted from the horn
in the cryostat.
We are also building a small cryogenic rotateable half-wave plate that
can be placed between the horn and the crossed-Dragone optics,
allowing the polarization angle to be continuously adjusted without
having to open the cryostat.


\subsection{Cryogenic Blackbody}
\label{sec:cryogenic_blackbody}


A slab of beam-filling Eccosorb absorber, which is coated with etched
Teflon for impedance matching, serves as a cold diffuse blackbody
load, and it is mounted directly in front of the horn apertures
approximately 1~cm from the detector module.
The temperature of the Eccosorb load is controlled using a heater
resistor and a weak thermal link that is connected to either the 3~K
stage of the pulse tube cooler or the 1~K stage of the ADR.
By changing the temperature of this load we can measure the absolute
brightness temperature calibration, which is used to compute the NET.
%


\subsection{Measurements}
\label{sec:measurements}


The sources described above will allow us to extensively characterize
our detectors.
In particular, we will be able to measure:
(i) calibrated NEP and NET under loading levels spanning 0.1-100~pW,
which correspond to the loading expected for a wide range of
experiments, including space-based, balloon-borne and ground-based
telescopes,
(ii) calibrated spectral response, both across the desired bands, but
also including any undesired out-of-band response,
(iii) detector response times to pulses of millimeter wavelength
radiation at realistic sky loading levels, and
(iv) the response of the detectors versus polarization angle of the
incoming radiation, including both the co-polarization and
cross-polarization response.
All of these measurements will be directly compared to our design
simulations and performance forecasts, providing the essential
feedback needed to identify any issues to be corrected in subsequent
wafer fabrication runs.


\section{DISCUSSION}
\label{sec:discussion}


One of the primary goals of this project is to bring the functionality
of MKIDs for CMB studies in line with state-of-the-art TES
bolometers\cite{bicep2_inst_2014,suzuki2012,arnold2012,polarbear12,rostem_2014,benson2014}.
To date we have (i) designed the critical broadband microstrip-to-CPW
coupler, (ii) modified the existing Advanced ACTPol design for SOI,
(iii) fabricated MKID optimization chips, and (iv) developed several
of the critical fabrication steps.
We are currently dark testing the MKIDs and laying out our final array
design.
Prototype array fabrication will begin in the summer of 2016.

In the future, we plan on making the sensing element in the MKIDs out
of aluminum manganese instead of aluminum.
By adding manganese to the aluminum, the $T_c$ of the sensor decreases
in a controllable way\cite{deiker_2004}, which does two critical
things.
First and foremost, in our current 150 and 235~GHz spectral bands, the
photons are energetic enough to break multiple Cooper pairs in the
sensing element, so the detector noise will be suppressed below the
photon noise -- even for the low optical loads that are expected in a
space-like environment.
Second, a lower $T_c$ makes the detector technology sensitive to lower
frequencies, so this technology will open the door to low-frequency
($\sim$30~GHz) MKIDs in the future.


\acknowledgments
 
This project is supported by a grant from the National Science
Foundation (Award \#1509211, \#1509078, and \#1506074) and a
NASA/NESSF Fellowship for McCarrick.


\bibliography{bjohnson_spie_2016}   
\bibliographystyle{spiebib}   


\end{document}